# Transfer current in p-type graphene/MoS$_2$ heterostructures

Khoe Van Nguyen,[1,2,3] Shih-Yen Lin,[1,4] Yia-Chung Chang[1,5,*]

[1]Research Center for Applied Sciences, Academia Sinica, Taipei 115, Taiwan

[2]Molecular Science and Technology, Taiwan International Graduate Program, Academia Sinica, Taipei 115, Taiwan

[3]Department of Physics, National Central University, Chungli, 320 Taiwan

[4]Graduate Institute of Electronics Engineering, National Taiwan University, Taipei 10617, Taiwan

[5]Department of Physics, National Cheng-Kung University, Tainan 701, Taiwan

**Abstract**

Transfer characteristics of p-doped graphene/monolayer-MoS$_2$ heterostructure at 300 K are measured experimentally and analyzed based on a model calculation. In the model, we first discretize the Poisson equation (PE) into multiple zones. In each zone the charge density is assumed constant and the PE can be transformed into a linear equation to determine the chemical potential self-consistently. To calculate the electrical conductivity, we solve the Boltzmann transport equation in the relaxation-time approximation using the inelastic acoustic phonon scattering by solving the Dirac equation. The relationship between the Dirac voltage (the voltage at which the Fermi level is located at the Dirac point) and the Fermi energy (i.e. the initial chemical potential at 0K) is derived. These calculations are performed without and with optical pumping and the results obtained agree well with the experimental data.

[*]**Corresponding author**. E-mail: yiachang@gate.sinica.edu.tw; Phone: +886 2 2787 3129

# 1. Introduction

Since graphene was fabricated in 2004 [1], there has been a huge interest in exploring properties of graphene due to its tremendous scientific and technological potential for applications in electronic and optical devices. The electronic band structures around the Dirac points can be well described by the Dirac cones, where the dynamics of charge carriers can be formulated by the relativistic Dirac-like equation with $v_F \approx c/300$ [2]. The total degeneracy of graphene is 4 [2]. Graphene intrinsically sustains high charge-carrier densities and possesses excellent room-temperature (RT) mobility but with low on/off ratio [1,2]. Different scattering mechanisms in graphene have been studied extensively both theoretically and experimentally [3-5]. For realistic applications, theoretical and experimental studies on optical and dielectric properties of graphene have also been investigated considerably [6-12]. The photo-absorbance of graphene is $\sim 2.3 \pm 0.1\%$ in the optical range and increases with increasing number of graphene monolayers [7]. The corresponding absorption coefficient is comparable to transition-metal dichalcogenices (TMDs). The out-of-plane dielectric constant of multi-layer (ML) graphene are field dependent, which approaches a constant at very low field, and at finite electric field the value of the dielectric constant is proportional to the ML-graphene thickness [9]. The microscopic out-of-plane dielectric permittivity of graphene in the hBN-graphene-hBN heterostructures at very low field is about $6.9\epsilon_0$ [11,12] with the effective dielectric thickness of graphene being about 0.22 nm [11]. Note that the graphene's thickness in van der Waals heterostructures is 0.341 nm [9]. For undoped (intrinsic) graphene, because of the zero gap, the work function ($W_G^0$) is equal to the electron affinity ($\chi_G$), i.e. $W_G^0 = \chi_G \approx 4.56$ eV [13-17]. While the electron affinity $\chi_G$ stays unchanged, the work function $W_G$ decreases with increasing gate voltage $\varphi_g$ [13]. Theoretical [17,19] and experimental [18,20] studies have confirmed that graphene is p-doped when contacting by gold (Au) electrodes with a Fermi energy in the range of 150-200 meV below the Dirac point, while still preserving its unique electronic structure. In addition, properties of interfacial adhesion between graphene and silicon dioxide ($SiO_2$) [21] substrate have also attracted much attention. In realistic graphene samples, there must exist many unavoidable defects; therefore, quantifying defect densities [22,23] plays a vital role in designing graphene-based devices. For point-like defects, the defect density can be calculated as $n_D = 1/\pi L_D^2$ [22,23] with $L_D$ being the average distance between defects.

Because graphene does not have a finite gap [2], its potential applications in electronic devices are clearly limited. In search of next-generation materials for nanoelectronics and optoelectronics, TMDs have emerged as attractive candidates [24-29]. Among these TMDs, 2H-MoS$_2$ has received increasing attention owing to its promising properties in optics, electronics, and optoelectronics [24-35]. Bulk MoS$_2$ and ML-MoS$_2$ films are semiconductors with indirect band gap of about 1.2 eV, whereas a SL-MoS$_2$ has a direct band gap of about 1.9 eV [24]. MoS$_2$ has much lower carrier mobility than graphene; however, MoS$_2$ not only has a higher on/off ratio but also a higher monolayer-absorption coefficient than graphene [24,28,32]. SL-MoS$_2$ has an absorbance of about 8% for free standing and about 5% on fused silica at the incident photon wavelength of 650 nm [28]. Therefore, MoS$_2$ can be used as an active region for creating photo-excited electron-hole pairs in MoS$_2$-based devices. The electronic band structures of MoS$_2$ near the band edges are parabolic [24,31] with a CB effective mass of 0.43-0.48 $m_e$ for SL-MoS$_2$ [31]. Besides the spin degeneracy of 2, ML- and SL-MoS$_2$ also exhibit CB valley degeneracies of 6 and 2, raising the total degeneracies to 12 and 4, respectively [24]. DFT calculations show that the electron affinity energy of SL-MoS$_2$ $\chi_M$ varies from 4.19 to 4.35 eV [16, 25, 26], which all fall in the range of the experimentally deduced value of $4.3 \pm 0.15$ eV for the bulk [27]. To the best of our knowledge, an experimental value of electron affinity of SL-MoS$_2$ is not available. The out-of-plane static dielectric properties of TMDs have been exhaustively studied from monolayer to bulk in [29]. It is worthy to note that, similar to graphene, the out-of-plane dielectric constant of ML-MoS$_2$ is field dependent, which seems to converge to a single value at a very low field, and at finite electric field the value of the dielectric constant is proportional to the ML-MoS$_2$ thickness [30]. Similarly, the work function $W_M$ decreases with increasing $\varphi_g$ for $\varphi_g$ above a threshold voltage $\varphi_T$ [32]. Theoretical [33-35,37] and experimental [34-36] investigations suggest that the presence of Mo and S vacancy defects can induce p-type and n-type midgap states, respectively, in pristine MoS$_2$; and thus both types of vacancy defects can coexist within one MoS$_2$ monolayer, where some defects serve as donors while others as acceptors, leading to electron localization [36]. Moreover, the existence of many types of Mo and S adatoms was also possible [34,35] and the vacancy concentration could be in the range of $3.5 \times 10^{13}/cm^2$ [35] or $9.5 \times 10^{12}/cm^2$ [37].

The above properties of graphene (G) and MoS$_2$ (M) are expected to complement each other in their van der Waals heterostructures (GMHs) such that one can design many useful devices. Therefore, GMHs are promising candidates for next-generation electronic and optoelectronic devices. Many experimental and theoretical studies of GMHs with various architectures have been reported in the literature [36,38-49]. Through GMHs, defect-induced midgap states in MoS$_2$ was probed and the experimentally estimated conduction band offset between graphene and trilayer-MoS$_2$ was about 170 meV [36]. In their vertical heterostructures, generation of highly efficient gate-tunable photocurrent was reported [38]; and an extraction of unusually efficient photocurrent by tunneling through discretized barriers was observed [45]. Under light illumination, the left shift of transfer conductivities and transfer currents in the negative direction of $\varphi_g$ due to the electron transfer from MoS$_2$ to graphene inducing the left shift of the Dirac points was recorded [39,42,44,48]. In the reference [40], the behavior of transfer conductivities of GMHs at different temperatures was investigated. The field-effect Schottky barrier transistors based on GMHs combining the qualities of high mobility from graphene and high on-off ratio from MoS$_2$ were made possible for the first time [41]. Recently, the authors in the reference [46] shed light on the charge transfer dynamics pathways in graphene/SL-MoS$_2$ heterostructures with the ultrafast electron transfer times on the femtosecond and sub-femtosecond scales. Now it seems that the use of GMHs for photosensing applications such as photodetectors may be on the way [38,44,45,48,49]. Finally, the crucial point in modeling these graphene/SL- and ML-MoS$_2$ is that the Dirac behavior of charge quasiparticles in graphene preserves [43,47]; thus, the dynamics of charge quasiparticles can be well described by the relativistic Dirac-like equation.

However, a quantitative device model to interpret experimental data and to predict and calculate relevant physical properties and quantities in GMHs is still lacking. In this manuscript, we report such a device model to explain the available experimental results for the electrical conductivity and transfer current in our as-prepared p-type GMH under a wide range of applied gate voltage at 300 K (RT) without and with optical pumping. We also establish the relationship between the Dirac voltage (the voltage at which the Fermi level is located at the Dirac point) and the Fermi energy (i.e. the initial chemical potential at 0K) induced by initial doping effects. Our device model can be extended to investigate n-type GMHs consisting of graphene and ML-MoS$_2$ including many-body effects under a wide range of applied gate voltage at various temperatures without and with optical pumping.

## 2. Material

In a previous study, one of us (SYL) successfully fabricated uniform large-size GMHs on sapphire substrates with layer-number controllability by chemical vapor deposition (CVD) [42]. Since we used the CVD chamber to sequentially grow the samples, our heterostructures (i.e. ML-M/G, SL-M/G and G/SL-M/G heterostructures on sapphire substrates) did not have any chemical contamination at the interfaces. The characteristics of these heterostructures were confirmed by the cross-sectional high-resolution transmission electron microscopy (HRTEM) image revealing the formation of GMHs, the X-ray photoelectron spectroscopy (XPS) spectra showing no chemical bonding between M and the underlying G, and the Raman spectrum demonstrating the uniform distribution of Raman frequency difference between $E_{2g}^1$ and $A_{1g}$ peaks across the film suggesting the controllability of the number of M monolayer(s) in the heterostructures. Moreover, the shift of the Dirac point in the G monolayer in the negative direction of the back-gate voltage induced by the photo-excited electrons being transferred from the M layer under light illumination also confirmed the formation of GMHs. A similar procedure of sample preparation and measurement is taken to fabricate a SL-M/G heterostructure on the SiO$_2$/Si substrate. Its architecture is illustrated in figure 1a.

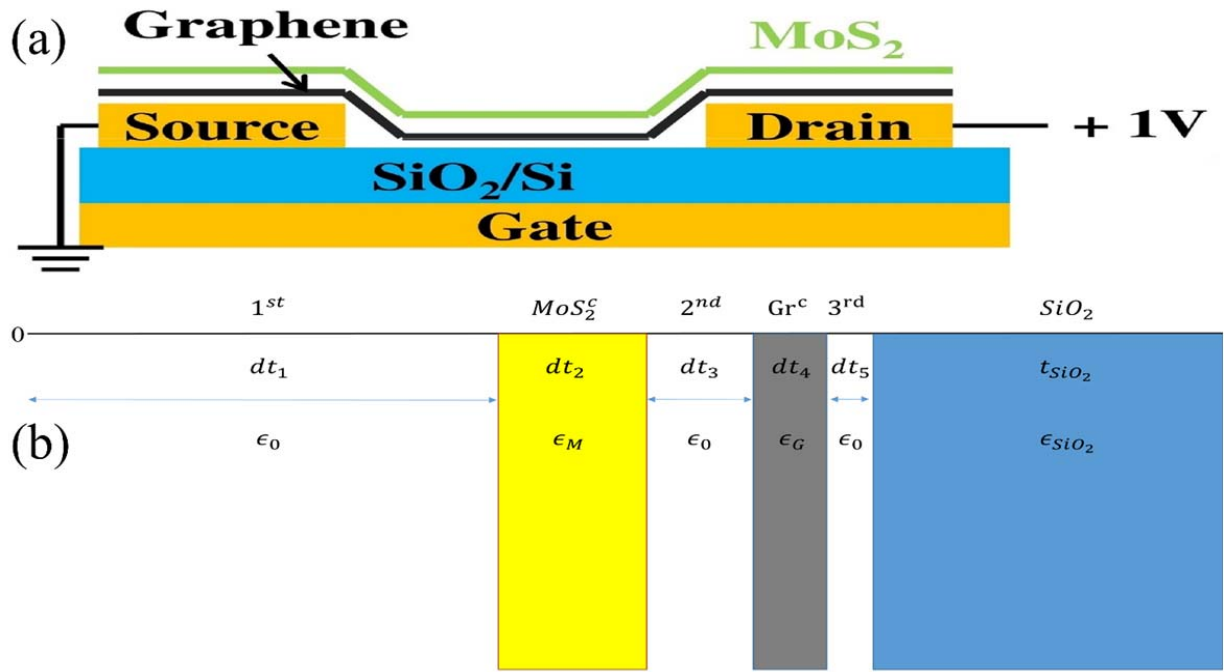

**Figure 1.** (a) The architecture of SL-MoS₂ and graphene (SL-M/G) on the SiO₂/Si substrate is considered in our experiment. The two contacts are put on the graphene monolayer. (b) The broadening model consists of three spacers named 1st, 2nd and 3rd and two core parts named $MoS_2^c$ and $Gr^c$ which are alternating on the SiO₂ substrate with a thickness of $t_{SiO_2}$ and a dielectric constant of $\epsilon_{SiO_2}$. $dt_i$ ($i = 1,2,...,5$) and $\epsilon_j$ ($j = 0, M, G$) are their effective thicknesses and dielectric constants, respectively. The schematically depicted relative ratios $dt_i/dt_j$ do not show the real relative ratios in the model. The positive $Oz$ is defined to be the direction from $O$ towards and perpendicular to the heterostructure.

## 3. Model

In our broadening model, we calculate the electrical conductivity for carrier transport in graphene by solving the Boltzmann transport equation in the relaxation-time approximation [50]

$$\sigma_0(T, \varphi_g) = \frac{e^2 v_F^2}{2} \int d\epsilon D_G(\epsilon) \tau_{tot}(T, \varphi_g; \epsilon) \left( -\frac{\partial f(\epsilon, \mu(\varphi_g); T)}{\partial \epsilon} \right),$$

where the total relaxation rate is defined by $\tau_{tot}^{-1}(T, \varphi_g; \epsilon) = \tau_{SR}^{-1} + \tau_{AP}^{-1}(T, \varphi_g; \epsilon) + \tau_{OP}^{-1}(T, \varphi_g; \epsilon) + \tau_{RIP}^{-1}(T, \varphi_g; \epsilon)$ with $\tau_{SR}^{-1}$ being considered to be the temperature- and energy-independent scattering rate due to the short-range scattering mechanisms overwhelmingly dominated by the defect scattering [3,4] and the relaxation rates $\tau_{AP}^{-1}(T, \varphi_g; \epsilon), \tau_{OP}^{-1}(T, \varphi_g; \epsilon), \tau_{RIP}^{-1}(T, \varphi_g; \epsilon)$ being due to the inelastic acoustic (AP) [5], optical (OP) and remote interfacial (RIP) [4] phonon scattering, respectively. The total contribution of $\tau_{AP}^{-1}(T, \varphi_g; \epsilon) + \tau_{OP}^{-1}(T, \varphi_g; \epsilon) + \tau_{RIP}^{-1}(T, \varphi_g; \epsilon)$ is much smaller than $\tau_{SR}^{-1}$, i.e. just about 3-5% as we will see later. Because scattering with AP has a greater contribution to the total scattering rate than scattering with OP and RIP for graphene on SiO₂ at RT [4], we will combine these two mechanisms into the short-range term for simplicity. The density of states of graphene is given by $D_G(\epsilon) = \frac{g_s g_v |\epsilon|}{2\pi \hbar^2 v_F^2}$ and $f(\epsilon, \mu(\varphi_g); T) = \left( e^{\frac{\epsilon - \mu(\varphi_g)}{k_B T}} + 1 \right)^{-1}$ denotes the Fermi-Dirac distribution with $\mu(\varphi_g)$ being the chemical potential, which approximately depends only on the applied gate voltage $\varphi_g$ and is determined self-consistently by solving the Poisson equation inside the multilayer system including SL-M/G on a SiO₂ substrate. Dimensionless quantities $g_s$ and $g_v$ are the spin and valley degeneracies, respectively, and $g_s = g_v = 2$ for graphene and a monolayer of

MoS$_2$. The low-energy band structures of graphene are described according to the Dirac equation $\epsilon_k = sgn(\epsilon_k)\hbar v_F |\vec{k}|$, where $sgn(\epsilon_k) = +1$ if $\epsilon_k \geq 0$ and $sgn(\epsilon_k) = -1$ if $\epsilon_k < 0$ corresponding to the conduction and valence bands, respectively. Due to the unavoidably inhomogeneous distribution of the gate voltage, by applying a normal distribution for $\varphi_g$, the electrical conductivity in graphene can be finally calculated to be

$$\sigma(T, \bar{\varphi}_g) = \frac{1}{\sqrt{2\sigma_g^2 \pi}} \int d\varphi_g \sigma_0(T, \varphi_g) \exp\left(-\frac{(\varphi_g - \bar{\varphi}_g)^2}{2\sigma_g^2}\right),$$

where $\bar{\varphi}_g$ is the mean or expectation of the distribution, $\sigma_g$ is the standard deviation and $\sigma_g^2$ is the variance of the gate voltage $\varphi_g$. The distribution of the gate voltage is completely homogeneous or, say, no broadening if $\sigma_g \approx 0$; in this critical limit, because the probability density function of the normal distribution becomes the Dirac delta function, we can rewrite the equation for the electrical conductivity in graphene as follows

$$\lim_{\sigma_g \to 0} \sigma(T, \bar{\varphi}_g) = \int d\varphi_g \sigma_0(T, \varphi_g) \delta(\varphi_g - \bar{\varphi}_g) = \sigma_0(T, \bar{\varphi}_g).$$

In general, the greater the value of $\sigma_g$ is, the more inhomogeneous the distribution of the gate voltage is or the more broadening. The transfer current can now be calculated from $I_{DS}(T, \bar{\varphi}_g; \varphi_{DS}) = \sigma(T, \bar{\varphi}_g) \frac{W}{L} \varphi_{DS}$, where $W$ and $L$ are the width and length of the channel, respectively, and $\varphi_{DS}$ is the drain-source bias. Now we move on to determine the chemical potential $\mu(\varphi_g)$ by using a newly developed matrix-based approach, which is described in details below, from which we can determine the Fermi-Dirac distribution $f(\epsilon, \mu(\varphi_g); T)$ and consequently calculate the carrier densities $n_\gamma(T, \varphi_g)$, the relaxation rate $\tau_{AC}^{-1}(T, \varphi_g; \epsilon)$ and, accordingly, the conductivity $\sigma_0(T, \varphi_g)$.

### 3.1 Poisson's equation

The electrostatic potential energy $V(\vec{r})$ of one point charge $q_e$ at $\vec{r}$ in the presence of an electrostatic potential $\varphi(\vec{r})$ is defined to be $V(\vec{r}) = q_e \varphi(\vec{r})$, where $q_e = -e$ for electrons. For simplicity, we drop the position notation $\vec{r}$ in equations without changing their meanings. Since the electric field is the gradient of the electrostatic potential, i.e. $\vec{E} = -\nabla\varphi$, applying Gauss's law of electrostatics results $\nabla \cdot \vec{E} = -\nabla \cdot \nabla\varphi = -\nabla^2\varphi = \frac{\rho_{3D}^{tot}}{\epsilon_0\epsilon_r}$. We thus obtain

$$\nabla^2\varphi = -\frac{\rho_{3D}^{tot}}{\epsilon_0\epsilon_r} = -\frac{q_e n_{3D}^{tot}}{\epsilon_0\epsilon_r} \Rightarrow \nabla^2 V = -\frac{q_e^2 n_{3D}^{tot}}{\epsilon_0\epsilon_r} = -\frac{q_e^2 n_{2D}^{tot}}{\epsilon_0\epsilon_r z},$$

where, by our conventions, $n_{2D/3D}^{tot} < 0$ for holes and $n_{2D/3D}^{tot} > 0$ for electrons.

### 3.2 The electrostatic potential energy in the broadening model

In our model, the spacers and the core parts of MoS$_2$ and graphene are alternate as shown in figure 1b. Because there is no charge in the spacers, we use a linear electrostatic potential of a general form of $V_s = b_s z + c_s$. Meanwhile, uniformly constant carrier densities in the core parts of MoS$_2$ and graphene are adopted, i.e. the electrostatic potentials have a quadratic form of $V_c = a_c z^2 + b_c z + c_c$. Therefore, we get general curvature

$$\nabla^2 V_c = 2a_c = -\frac{q_e^2 n_{2D}^{tot}}{\epsilon_0\epsilon_r z_c},$$

where $\nabla^2 V_c > 0$ (i.e. the electrostatic potential is concave) for a hole density and $\nabla^2 V_c < 0$ (i.e. the electrostatic potential is convex) for an electron density. The averaged electrostatic potential in the core parts of MoS$_2$ and graphene is generally defined to be

$$\bar{V}_c = \frac{1}{z_c}\int_0^{z_c} V_c \, dz = \frac{1}{3}a_c z_c^2 + \frac{1}{2}b_c z_c + c_c.$$

Note that, because the band gap of SiO$_2$ is too huge for carriers to transfer, we assume a linear electrostatic potential for this substrate just like the spacers, i.e. $V_{SiO_2} = b_{SiO_2} t_{SiO_2} + c_{SiO_2}$.

### 3.3 Calculations of 2D carrier densities

We follow a standard approach to calculate free carrier densities in the monolayers as follows

$$n_\gamma(T,\varphi_g) = \int d\epsilon D_\gamma f_\gamma(\epsilon, \mu_\gamma(\varphi_g); T),$$

where $\gamma = G, M$ stands for graphene and MoS$_2$, respectively. $D_\gamma$ is the density of states of $\gamma$ and $f_\gamma(\epsilon, \mu_\gamma(\varphi_g); T)$ is the Fermi-Dirac distribution of $\gamma$. In general, $\mu_\gamma(\varphi_g) = \mu_G^0 - \Delta_\gamma - \bar{V}_\gamma(\varphi_g)$. Here $\mu_G^0$ is the initial Fermi level, i.e. the Fermi level at zero gate voltage induced by the initial doping $\mu_G^0 = \mu_G(0)$, $\bar{V}_\gamma(\varphi_g)$ being the averaged electrostatic potential of $\gamma$ such that in graphene for instance $\bar{V}_G(0) = 0$ and $\bar{V}_G(\varphi_D) = \mu_G^0$ at the Dirac voltage $\varphi_D$ – the gate voltage at which the Fermi level is located at the Dirac points $\mu_G(\varphi_D)=0$, $\Delta_\gamma$ being the conduction band offset between graphene and MoS$_2$, i.e. $\Delta_G = 0$ and $\Delta_M = CBM_M - CNP_G \equiv \Delta$ with $CBM_M$ the conduction band minimum in MoS$_2$ and $CNP_G$ the charge neutrality point (i.e. the Dirac point) in graphene which is, by usual convention, chosen to be the energy origin.

Then, before carriers in graphene transfer into MoS$_2$, the total carrier density in the channel is the sum of the free carrier density $n_G(T, \varphi_g)$, which consists of the free carrier density at zero gate voltage $n_G^0 = n_G(T, 0)$ and the free carrier density induced by finite gate voltage $n_G^{non-0} = n_G(T, \varphi_g \neq 0)$, and the dopant density $n_{dp}$ in the channel. That is

$$n_G^{tot}(T, \varphi_g) = n_G(T, \varphi_g) + n_{dp} = n_G^0 + n_G^{non-0} + n_{dp}.$$

Note that $n_G^0 + n_{dp} = 0$ and $n_G^{non-0}(T, 0) = 0$. This means that the total carrier density in the channel is exactly equal to the free carrier density induced by the non-zero gate-voltage term, i.e. $n_G^{tot}(T, \varphi_g) = n_G^{non-0}$ despite of the initial doping.

**3.4 Determination of the Fermi level based on a newly developed approach**

In our broadening model, because of the *ab initio* fact that the dielectric constant of graphene is not a true constant but decreases rather fast to the vacuum permittivity when approaching the two opposite surfaces of the graphene monolayer [11,12], we apply the same idea to MoS$_2$ and, as a result, we adopt an additional spacer between the two successive core parts of the two successive monolayers as indicated in the figure 1b. Note that h-BN also displays such dielectric behavior [11,12]. Because the electrostatic potentials and displacement field must be continuous at the boundaries between the spacers and the core parts, we can establish a system of coupled equations depending only on the gate voltage $\varphi_g$ and the distant $dt_1$ from the surface of the core part of the first monolayer, MoS$_2$ in our GMH for instance, outwards to the half of the thickness of the Au contacts, where $c_0$ is between 0 and $-\varphi_{DS}$ with $\varphi_{DS}$ being the drain-source voltage. If a heterostructure has $n_{ml}$ monolayers of the same or different materials excluding the substrate, SiO$_2$ in our GMH for instance, due to an additional spacer between the two successive core parts of the two successive monolayers, the total number of the core parts is $n_{cp} = n_{ml}$ and the total number of the spacers equals $n_{sp} = n_{cp} + 1 = n_{ml} + 1$; therefore, the total number of interfaces is $n_{if} = 2n_{ml} + 1$, where plus 1 is due to the last interface between the last spacer and the substrate. Accordingly, the total number of equations being equal to the total number of unknowns is $n_{eq} = 2n_{if} + 1$, where the last equation relates the potential in the substrate and the gate voltage $\varphi_g$ and two equations are used at each interface to describe the boundary conditions for electrostatic potentials and their derivatives. As a result, $n_{eq} = 4n_{ml} + 3$. Our GMH in the figure 1a has three spacers and two core parts being alternate, i.e. the first spacer/the core part of MoS$_2$/the second spacer/the core part of graphene/the third spacer on the SiO$_2$ substrate, as shown in the figure 1b and thus $n_{ml} = 2 \rightarrow n_{if} = 5 \Rightarrow n_{eq} = 11$ equations as follows $c_0 + b_1 dt_1 = c_1$, $c_1 + b_2 dt_2 + a_M dt_2^2 = c_2$, $c_2 + b_3 dt_3 = c_3$, $c_3 + b_4 dt_4 + a_G dt_4^2 = c_4$, $c_4 + b_5 dt_5 = c_5$, $c_5 + b_{SiO_2} t_{SiO_2} = V_g$, $b_1 = \epsilon_M b_2$, $\epsilon_M (2a_M dt_2 + b_2) = b_3$, $b_3 = \epsilon_G b_4$, $\epsilon_G (2a_G dt_4 + b_4) = b_5$, $b_5 = \epsilon_{SiO_2} b_{SiO_2}$. Here $V_g \equiv q_e \varphi_g$ and $f_{cM}$ and $f_{cG}$ are the fractions of the core parts of MoS$_2$ and graphene, respectively; $t_M$, $t_G$ and $t_{SiO_2}$ are the thicknesses of MoS$_2$, graphene monolayers and of SiO$_2$, respectively. After solving this system of coupled equations self-consistently by implementing the Poisson equation above at each gate voltage $\varphi_g$ at RT, we obtain $\bar{V}_\gamma(\varphi_g)$. Substitute $\bar{V}_\gamma(\varphi_g)$ into $\mu_\gamma(\varphi_g) = \mu_G^0 - \Delta_\gamma - \bar{V}_\gamma(\varphi_g)$ of $f_\gamma(\epsilon, \mu_\gamma(\varphi_g); T)$, we can calculate $n_\gamma(T, \varphi_g)$ and $\sigma_0(T, \varphi_g)$ without and with optical pumping. Thus, for each $c_0 = -e\varphi(x) = -e\varphi_{DS} x/L \in [0, -\varphi_{DS}]$ eV with $x \in [0, L]$, we determine all physical quantities. Then we take the arithmetic mean to obtain the average results. Here it is reasonable to assume that the system is homogeneous or has a translational symmetry in the transverse (perpendicular) y-direction (i.e. the channel width direction $W$) to the drain-source x-direction (i.e. the channel length direction $L$).

## 3.5 Relationship between the Dirac voltage and the initial chemical potential at 0K

Because of the fact that $f_\gamma(\epsilon, \mu_\gamma(\varphi_g); 0) = 1$ and quantum capacitance contributions in a series are negligible [40], the total free carrier density induced by the oxide capacitance is given by [40]

$$\frac{sgn(n_G) c_G \left(\mu_G^0 - \bar{V}_G(\varphi_g)\right)^2}{2} + n_M(0, \varphi_g) = \frac{\epsilon_0 \epsilon_{ox}}{e t_{ox}} (\varphi_g - \varphi_D).$$

Here $c_G = \frac{g_s g_v}{2\pi \hbar^2 v_F^2}$ is a prefactor in the density of states of graphene $D_G(\epsilon) = \frac{g_s g_v |\epsilon|}{2\pi \hbar^2 v_F^2}$. For not heavily doped graphene monolayers such that there is no carrier transfer into MoS$_2$ at zero gate

voltage, i.e. $n_M(0,0) = 0$, since $\bar{V}_G(0) = 0$, we get the relationship between the initial doping $\mu_G^0$ and the Dirac voltage $\varphi_D$ as $\mu_G^0 = \pm\sqrt{\frac{-2\epsilon_0\epsilon_{ox}sgn(n_G^0)\varphi_D}{et_{ox}c_G}}$. Then we resubstitute this equation for $\mu_G^0$ into the equation for the total free carrier density induced by the oxide capacitance and reasonably assume that graphene monolayers are not heavily doped, we obtain the following equation for the chemical potential in graphene at any $\varphi_g$

$$\mu_G(\varphi_g) = \mu_G^0 - \bar{V}_G(\varphi_g) = \pm\sqrt{\frac{2\epsilon_0\epsilon_{ox}sgn(n_G)(\varphi_g-\varphi_D)}{et_{ox}c_G}},$$

from which we can determine the averaged potential at any $\varphi_g$ in graphene $\bar{V}_G(\varphi_g)$, the carrier density in graphene $n_G(0, \varphi_g)$ and the electrical conductivity in graphene $\sigma_0(0, \varphi_g)$.

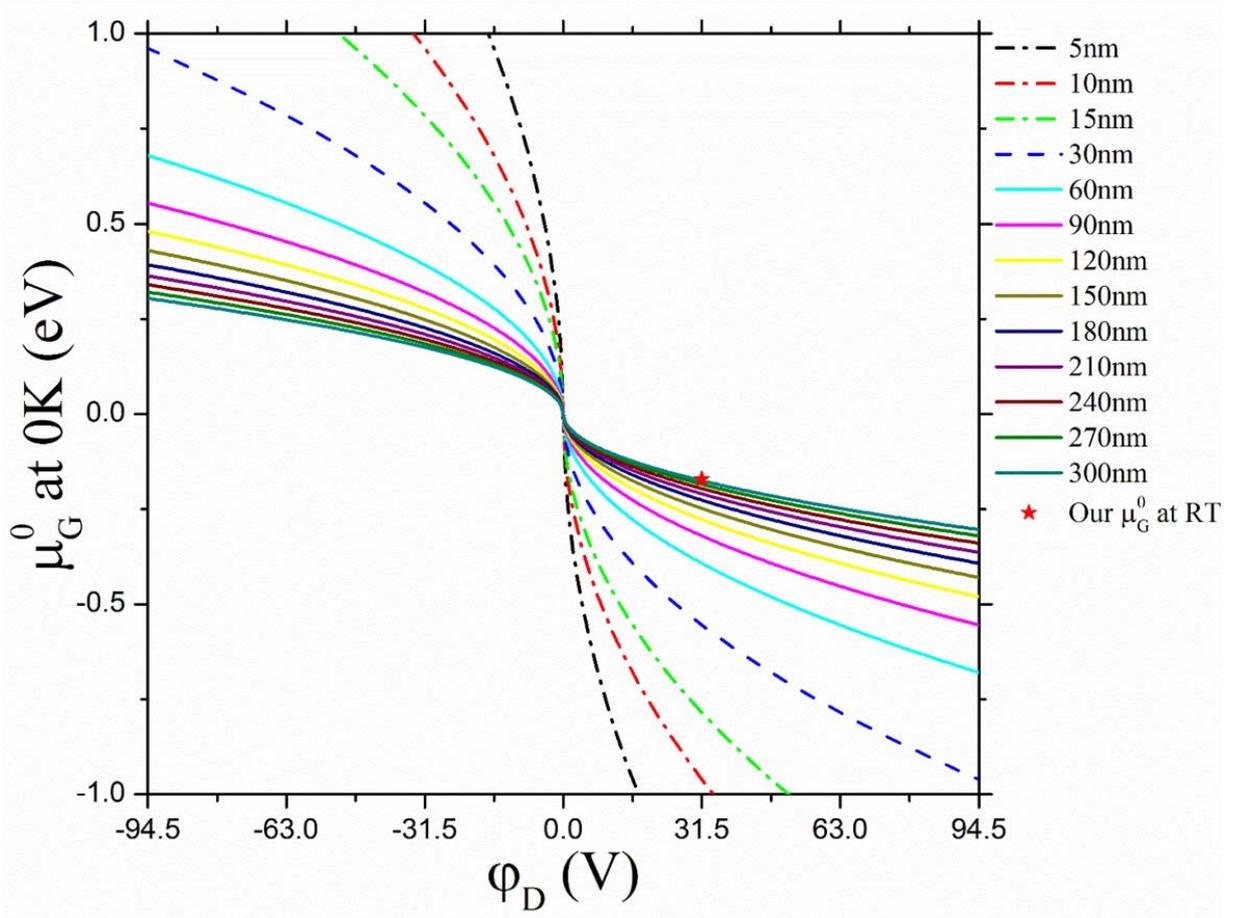

**Figure 2.** The relationship between the initial doping $\mu_G^0$ and the Dirac voltage $\varphi_D$ for different values of the thickness of $SiO_2$ at 0K. The red star is $\mu_G^0 = -171$ meV we use to model our experimental data for graphene encapsulated between SL-MoS$_2$ and 300-nm SiO$_2$ at RT. As an effect of temperature, our $\mu_G^0$ is 4.4 meV greater than the calculated value of $-175.4$ meV.

Note that, before carriers in graphene transfer into MoS$_2$ at any $T$ and $\varphi_g$, we have $n_G(T, \varphi_g) = \frac{\epsilon_0 \epsilon_{ox}}{e t_{ox}} (\varphi_g - \varphi_D) \Rightarrow n_G^0 = -\frac{\epsilon_0 \epsilon_{ox}}{e t_{ox}} \varphi_D$ and $n_G^{non-0} = \frac{\epsilon_0 \epsilon_{ox}}{e t_{ox}} \varphi_g$ satisfying that $n_G^{non-0}(T, 0) = 0$ implying $n_G^{tot}(T, \varphi_g) = n_G(T, \varphi_g) + n_{dp} = n_G^0 + n_G^{non-0} + n_{dp} = n_G^{non-0} = \frac{\epsilon_0 \epsilon_{ox}}{e t_{ox}} \varphi_g$.

Figure 2 shows that, if $t_{SiO_2} < 30\ nm$, $|\mu_G^0|$ is very sensitive to $\varphi_D$ and may exceed 1 eV. If $t_{SiO_2}$ satisfies $30\ nm \leq t_{SiO_2} < 120\ nm$, $0.5\ eV \leq |\mu_G^0| \leq 1\ eV$ for $30\ V \ll |\varphi_D| < 100\ V$. And if $t_{SiO_2} \geq 120\ nm$, $|\mu_G^0| \leq 0.5\ eV$ for $|\varphi_D| \leq 100\ V$. This is the range of interest that practical electronic devices often work. In our set-up, $t_{SiO_2} = 300\ nm$ and $\varphi_D = 31.5\ V$ lead to $\mu_G^0 \approx -0.175\ eV$ which is slightly smaller than our initial chemical potential (i.e. the red star) needed to fit the data at RT, i.e. $-175.4\ meV$ versus $-171\ meV$. We attribute this discrepancy to the temperature effect. Note that $\mu_G^0 = -171\ meV$ (between -200 and -150 meV) [17-20].

We can use the relationship between $\mu_G^0$ and $\varphi_D$ for graphene on other substrates such as hexagonal boron nitride (h-BN), silicon carbide (SiC), sapphire (Al$_2$O$_3$), hafnium dioxide (HfO$_2$),... Intrinsically, this relationship only depends on the thickness and the dielectric constant of the substrate. Thus, based on this relationship, we may engineer some electronic properties of graphene-based devices by choosing which material and how thick it is to make the substrate.

### 3.6 Computational details

The Fermi velocity is given by $v_F = \frac{c}{300}$. The constants $e, \hbar, c, k_B, T$ are the fundamental electric charge, reduced Planck constant, light velocity in vacuum, Boltzmann constant, and temperature, respectively. For graphene, a monolayer has a thickness $t_G = 0.341\ nm$ [9], a static core (effective) dielectric constant $\epsilon_G = 6.9\epsilon_0$ [11,12] and a fraction of the core part $f_G \approx 0.645$ such that the effective (core) dielectric thickness of graphene is $f_G t_G = 0.22\ nm$ given by [11]. Because the thickness of a MoS$_2$ monolayer is $t_M = 0.670\ nm$ [30], we consider the core part of the MoS$_2$ monolayer as having an effective dielectric thickness of $f_M t_M = 0.612\ nm$ [29] and a corresponding static microscopic dielectric permittivity of $\epsilon_M = 6.4\epsilon_0$ [29]. Thus $f_M \approx 0.913$. Finally, the thickness of the substrate is $t_{SiO_2} = 300\ nm$ with the well-known dielectric constant

$\epsilon_{SiO_2} = 3.9\epsilon_0$. Note that, because our GMH has a large channel area with a channel width of $W = 10\ \mu m$ and a channel length of $L = 25\ \mu m$, we do not have to consider the finite-size effect despite of the small aspect ratio $W/L = 0.4$. The two Au contacts have a thickness of $t_{DS} = 50\ nm$ and the drain-source bias $\varphi_{DS} = 1\ V$, which is, except for a very narrow range of gate voltage around the Dirac voltage $\varphi_D$, much smaller than the applied voltage $\varphi_g \sim (-95, 95)\ V$. We take the drain-source bias's effects into account by computing all physical quantities at a sampling of $c_0$ in the interval of $[0, -1]\ eV$ at $dt_1 = \frac{t_{DS}}{2} - (\frac{f_M t_M}{2} + d_{C-M} + t_G)$, where $d_{C-M}$ is the distant from the C-plane to the Mo-plane, and then we take the arithmetic mean to obtain the average results. As we have already mentioned earlier, we have reasonably assumed that the system is homogeneous in the transverse y-direction (i.e. the channel width direction $W$) to the drain-source x-direction (i.e. the channel length direction $L$). Other parameters are $dt_2 = f_M t_M$, $dt_3 = d_{C-M} - (\frac{f_G t_G}{2} + \frac{f_M t_M}{2})$, $dt_4 = f_G t_G$, $dt_5 = t_G - \frac{f_G t_G}{2}$.

## 4. Results

### 4.1 Without optical pumping

Before carrier transfer, the monolayer of MoS₂ plays a role of a top dielectric layer. So the averaged potential, the chemical potential, the carrier density, and the electrical conductivity of graphene stay almost unchanged as when MoS₂ is removed from the GMH leaving only graphene on SiO₂. The carrier density transferred into MoS₂ can be calculated as $n_M(T, \varphi_g) = c_M\left(\mu_G^0 - \Delta - \bar{V}_M(\varphi_g)\right)$, where $c_M = \frac{g_s g_v m^*}{2\pi\hbar^2}$ is the density of states of MoS₂. It is well-known that $g_s = g_v = 2$ [24] for a monolayer of MoS₂; $m^* = 0.43 - 0.48\ m_e$ is the effective mass of electrons [31] with $m_e$ being the bare electron mass. Here the Fermi level is measured with respect to the CBM in MoS₂, whereas it is measured with respect to the Dirac point in graphene.

The averaged electrostatic potential and the chemical potential in graphene $\mu_G(\varphi_g) = \mu_G^0 - \bar{V}_G(\varphi_g)$ calculated without and with considering carrier transfer are plotted in Fig. 3a. Clearly, stably self-consistent solutions are found for both cases and $\mu_G(0) = \mu_G^0 = -171\ meV$ resulting $n_G^0 \approx -2.3 \times 10^{12}/cm^2$ and $\mu_G(31.5\ V) = \mu_G^0 - \bar{V}_G(31.5\ V) = 0$, i.e. $\varphi_D = 31.5\ V$. To fit our experimental data (i.e. the transfer current in Fig. 4), electrons in graphene must transfer into MoS₂ at a transfer gate voltage $\varphi_T \approx 44.5\ V$ as indicated by $\bar{V}_G(\varphi_T)$ and $\mu_G(\varphi_T)$ in Fig. 3a, which is translated into a transfer energy level of about 107 meV above the Dirac point in graphene. This value is reasonably just below the experimentally estimated conduction band offset between graphene and trilayer-MoS$_2$ of about 170 meV [36].

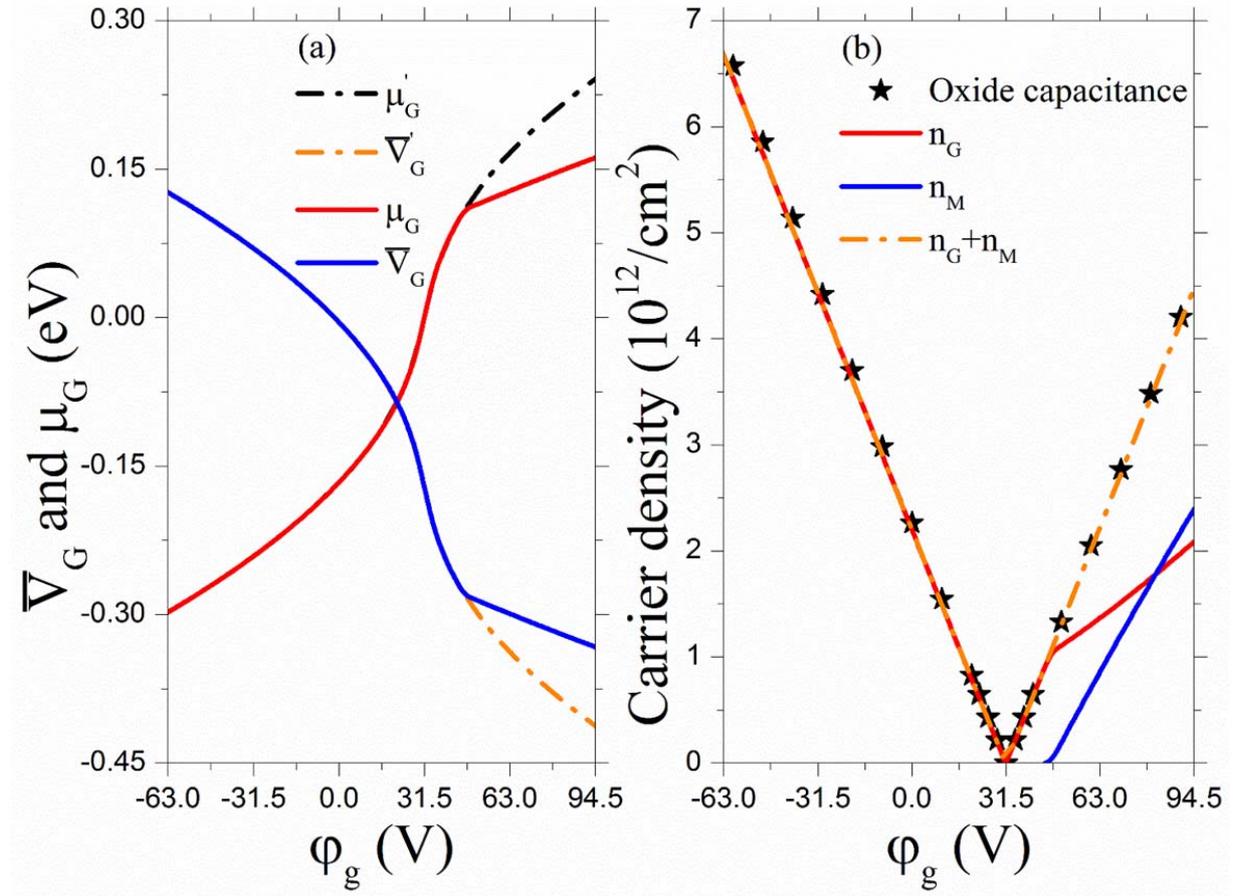

**Figure 3.** (a) The averaged electrostatic potential and the chemical potential of the as-grown p-doped monolayer of graphene encapsulated between a monolayer of MoS$_2$ and a substrate of SiO$_2$ as a function of the gate voltage are calculated without (the dash dot curves) and with (the solid curves) carrier transfer from graphene into MoS$_2$. (b) The carrier density as a function of the gate voltage. The stars display the total carrier density computed by using the oxide capacitance equation, which are slightly higher than the sum of carrier densities induced in MoS$_2$ and graphene by self-consistent calculations; the larger $|\varphi_g - \varphi_D|$, this deviation is bigger but it is in general negligible.

Fig. 3b demonstrates the carrier densities in the GMH. Obviously, carrier transfer starts at $\varphi_T \approx 44.5\ V$ and when $\varphi_g > 81\ V, n_M > n_G$ but $n_G + n_M \simeq n_{tot}$ calculated by the oxide capacitance equation [40]. Our slightly underestimation may be attributed to the quantum capacitane contributions, which have been ignored in our self-consistent calculations as we have already mentioned above.

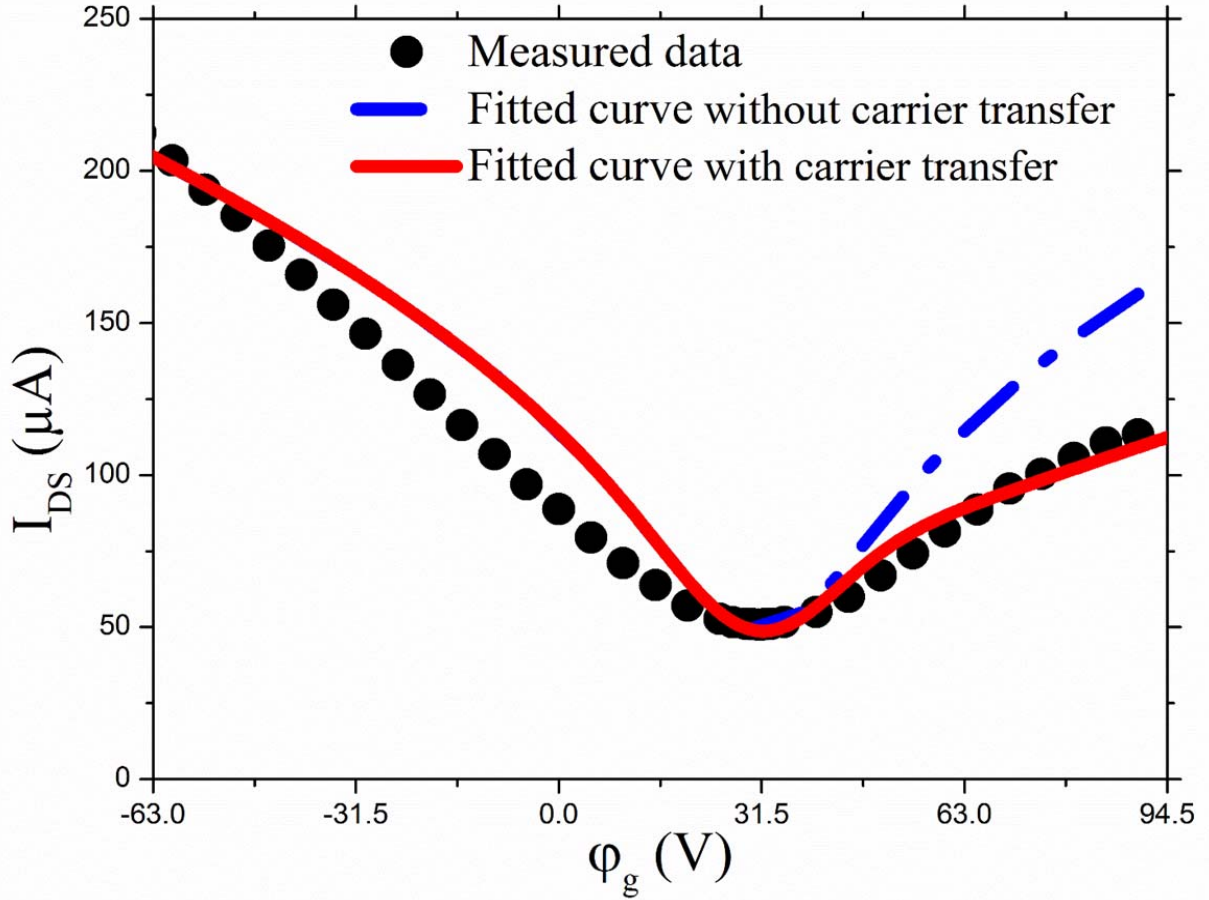

**Figure 4.** The transfer current as a function of the gate voltage. Both the transfer currents without (the dash dot curve) and with (the solid curve) carrier transfer are plotted. Note that, for simplicity, only carrier transfer when $\varphi_g > \varphi_D$ is considered.

The transfer current as a function of the gate voltage $I_{DS}(\varphi_g)$ is illustrated in Fig. 4. We see that the transfer current with carrier transfer agrees very well with the measured data. Here we note that, because we only use the carrier transfer for the right branch (i.e. when $\varphi_D < \varphi_g$, electrons are carriers; so the channel is called the n-channel or the electron-channel), the p-channel or the hole-channel (i.e. the left branch or when $\varphi_g < \varphi_D$) deviates a little bit from the experimental data. This channel can be improved by using a similar procedure.

**4.2 With optical pumping**

We have used light of wavelength of $650\ nm$ with the power density of $10\ \frac{mW}{cm^2}$ to illuminate our as-grown p-type GMH. Quantum mechanically, this light consists of photons of energy of about $1.9\ eV$. Therefore, after absorbing these photons, electrons in the VB gain enough energy and jump up into the CB leaving holes behind in the VB generating electron-hole pairs named excitons in the intrinsic monolayer of MoS$_2$. At zero gate voltage, because the CBM of intrinsic MoS$_2$ is some hundreds of $meV$ higher than the initial chemical potential of p-doped graphene (i.e. $\Delta - \mu_G^0 \sim 170 + 170 = 340\ meV$ [36]), these electrons almost immediately transfer into the p-doped graphene monolayer and deplete holes there. This is why the Dirac voltage shifts to the left compared to the case of no optical pumping [39,42,44,48]. The electron transfer times are ultrafast and on the femtosecond and sub-femtosecond scales [46].

After the system reaches a steady state, the Dirac voltage becomes $\varphi_D^{opt} \approx -4.5\ V$, the initial chemical potential being $\mu_G^{0-opt} \approx 35\ meV$, the initially free carrier density in graphene being $n_G^{0-opt} \approx 0.2 \times 10^{12}/cm^2$, the total carrier density in graphene (i.e. electrons transferred from the CB of MoS$_2$) being $n_G^{tot-opt} = -n_G^0 + n_G^{0-opt}$, and the total carrier density in MoS$_2$ (i.e. holes left in the VB of MoS$_2$) being $n_M^{tot-opt} = -n_G^{tot-opt} = n_G^0 - n_G^{0-opt}$. The p-type GMH now becomes a parallel plate capacitor, which has a totally negative charge density of $\rho_{G-e}^{tot-opt} = e(n_G^0 - n_G^{0-opt})$ in graphene and a totally positive charge density of $\rho_{M-h}^{tot-opt} = -\rho_{G-e}^{tot-opt} = -e(n_G^0 - n_G^{0-opt})$ in the VB of MoS$_2$ such that the system is neutral. Here we recall that $n_G^0 \approx -2.3 \times 10^{12}/cm^2 < 0$ due to the as-prepared p-doped graphene and $e > 0$ by our conventions.

Completely similar to the previous case for no optical pumping, the averaged electrostatic potential and the chemical potential in graphene $\mu_G(\varphi_g) = \mu_G^0 - \bar{V}_G(\varphi_g)$ calculated without and with considering carrier transfer are plotted in Fig. 5a. Clearly, we find stably self-consistent solutions for both cases and $\mu_G(0) = \mu_G^{0-opt} \approx 35\ meV$ resulting $n_G^{0-opt} \approx 0.2 \times 10^{12}/cm^2$ and

$\mu_G(-4.5\,V) = \mu_G^0 - \bar{V}_G(-4.5\,V) = 0$, i.e. $\varphi_D = \varphi_D^{opt} \approx -4.5\,V$. To fit our experimental data (i.e. the transfer current in Fig. 6), electrons in graphene must transfer into MoS$_2$ at a transfer gate voltage $\varphi_T \approx 7\,V$ as indicated by $\bar{V}_G(\varphi_T)$ and $\mu_G(\varphi_T)$ in Fig. 5a, which is translated into a transfer energy level of about 140 meV above the Dirac point in graphene. This value is reasonably just below the experimentally estimated conduction band offset between graphene and trilayer-MoS$_2$ of about 170 meV [36] and reasonably higher than the value of 107 meV in the case of no optical pumping. It comes from the fact that all/most of midgap levels of 107-139 meV may have been depleted and thus electrons may have been localized there [36] before optical pumping. Fig. 5b demonstrates the carrier densities in the GMH. Obviously, carrier transfer starts at $\varphi_T \approx 7\,V$ and when $\varphi_g > 26\,V, n_M > n_G$ but $n_G + n_M \simeq n_{tot}$ calculated by the oxide capacitance equation [40]. Our slightly underestimation may be attributed to the quantum capacitance contributions, which have been ignored in our self-consistent calculations as we have already mentioned above.

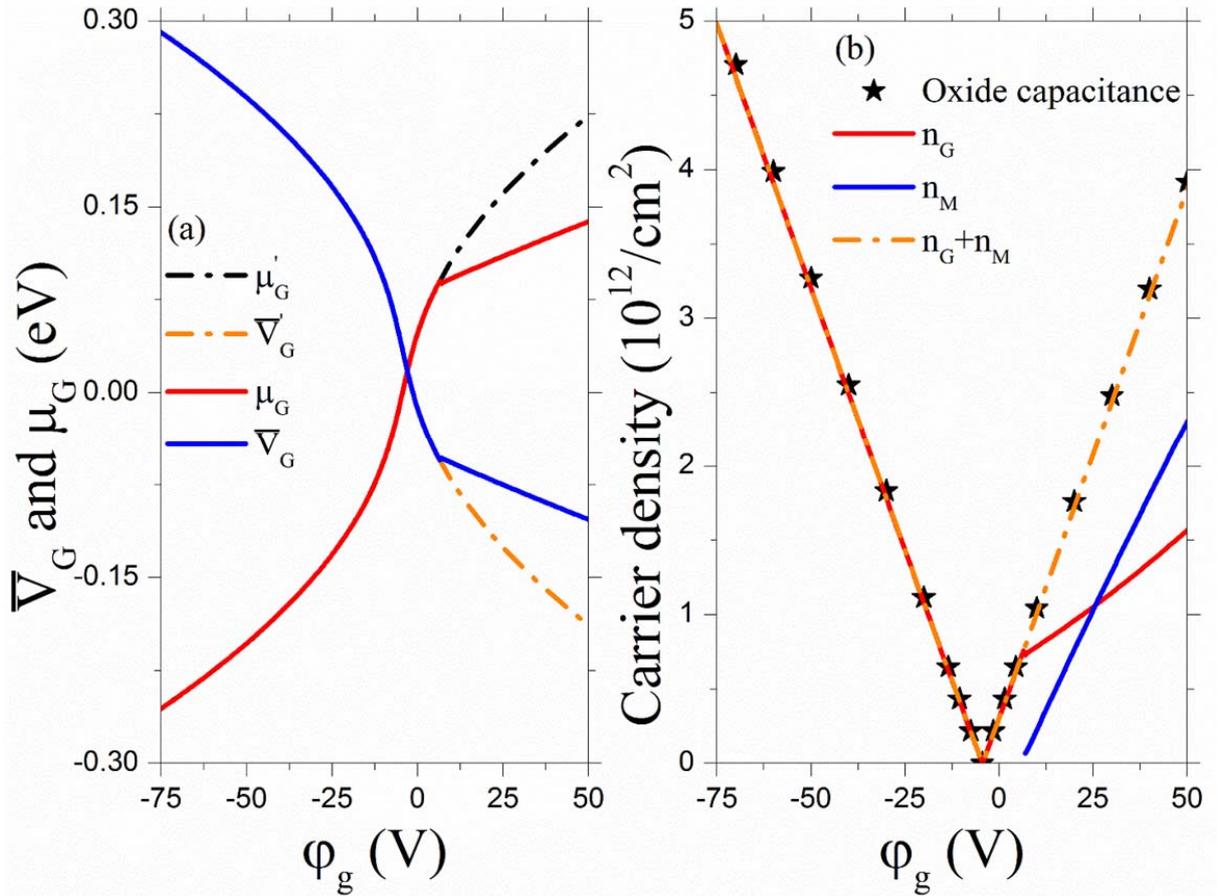

**Figure 5.** (a) The averaged electrostatic potential and the chemical potential of the as-grown p-doped monolayer of graphene encapsulated between a monolayer of MoS$_2$ and a substrate of SiO$_2$ as a function of the gate voltage are calculated without (the dash dot curves) and with (the solid curves) carrier transfer from graphene into MoS$_2$ under light illumination. (b) The carrier density as a function of the gate voltage under light illumination. The stars display the total carrier density computed by using the oxide capacitance equation, which are slightly higher than the sum of carrier densities induced in MoS$_2$ and graphene by self-consistent calculations; the larger $|\varphi_g - \varphi_D|$, this deviation is bigger but it is in general negligible.

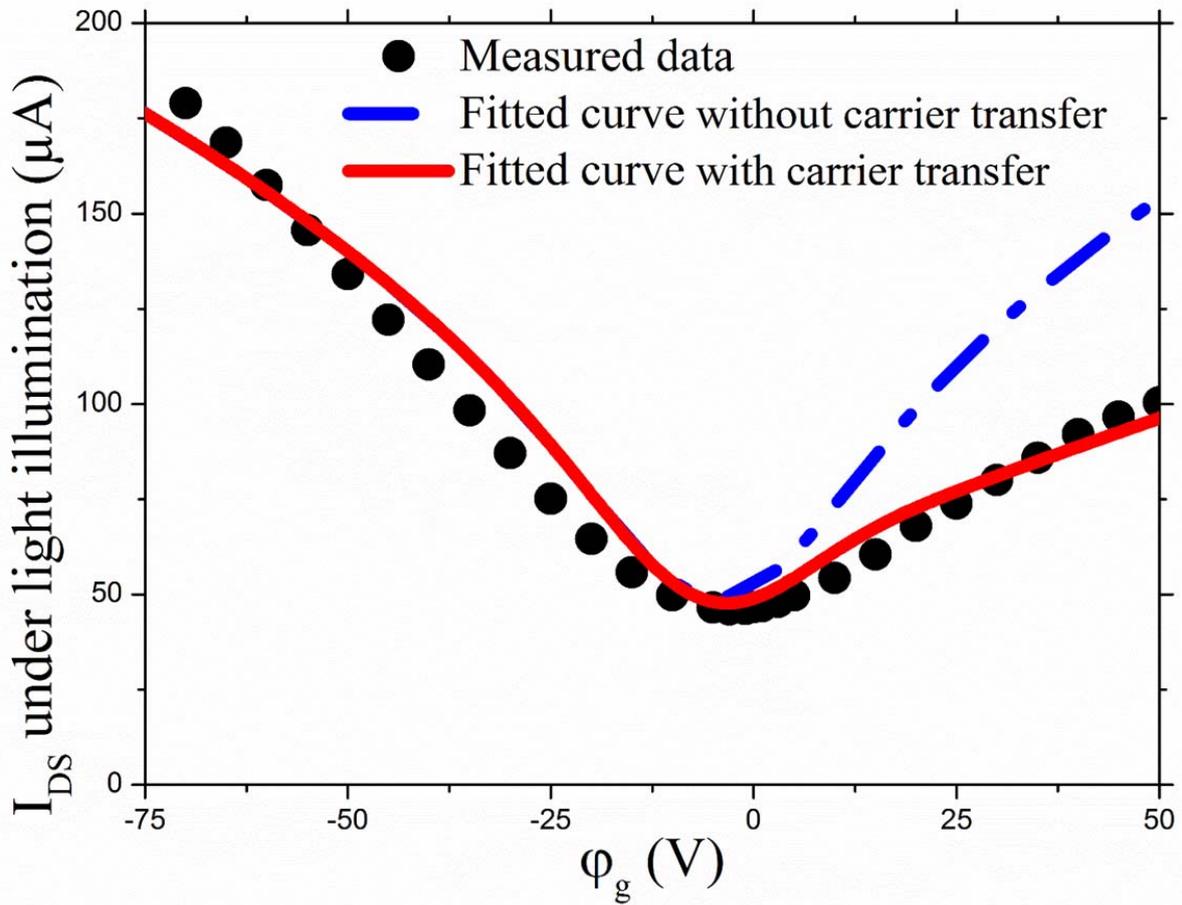

**Figure 6.** The transfer current as a function of the gate voltage under light illumination. Both the transfer currents without (the dash dot curve) and with (the solid curve) carrier transfer are plotted. Note that, for simplicity, only carrier transfer when $\varphi_g > \varphi_D$ is considered.

The transfer current as a function of the gate voltage $I_{DS}(\varphi_g)$ is illustrated in Fig. 6. We see that the transfer current with carrier transfer agrees very well with the measured data. Here we note that, because we only use the carrier transfer for the right branch (i.e. when $\varphi_D < \varphi_g$, electrons are carriers; so the channel is called the n-channel or the electron-channel), the p-channel or the hole-channel (i.e. the left branch or when $\varphi_g < \varphi_D$) deviates a little bit from the experimental data. This channel can be improved by using a similar procedure.

Note that we have already used two physically free parameters to fit the transfer currents in graphene, i.e. a standard deviation due to broadening $\sigma_g \approx 9\,V$ and $\hbar\tau_{SR}^{-1} \approx 44\,meV$ or $\tau_{SR}^{-1} \approx 66\,THz$, which falls in an experimental range of $\sim 2 - 100\,THz$ [3,4] and is overwhelmingly greater than $\tau_{AP}^{-1} < 2\,THz$ [4,5] for $|\mu_G| < 0.5\,eV$ at RT. Moreover, $\tau_{SR}^{-1} \approx 66\,THz$ is translated into $L_D \approx 15\,(nm)$ and thus a defect density $n_D = \frac{1}{\pi L_D^2} \approx 1.4 \times 10^{11}(cm^{-2})$, which falls in an experimental range of $\sim 4.20 \times 10^{10} - 1.38 \times 10^{13}/cm^2$ [22,23]. Finally, there always exists an inflection point in $\mu_G^0$ at $\varphi_D = 0$ (Fig. 2) and in $\bar{V}_G$ and $\mu_G$ at $\varphi_g = \varphi_D$ (Figs. 3a and 5a) as a direct consequence of the electron-hole symmetry described by the Dirac cones in graphene.

## 5. Summary and Conclusions

We have successfully conducted the theoretical self-consistent calculations on GMHs. We obtain reasonable results agreeing well with the experimental data on our as-prepared p-doped GMH at RT under a large range of the gate voltage without and with optical pumping. Physical quantities such as the averaged electrostatic potential, the chemical potential, the carrier densities, the electrical conductivity and the transfer current are reasonably computed in details and comparable to the available experimental data. In addition, we also establish an equation relating the initial Fermi energy to the Dirac voltage at 0K. This relationship helps us estimate one of them when knowing the other at finite temperatures within a high precision. Additionally, we may engineer desired electronic properties of graphene-based devices based on this equation by choosing appropriate materials characterized by appropriate dielectric constants and thicknesses. Finally, our broadening model can be extended to investigate n-type GMHs consisting of graphene and ML-MoS$_2$ and other graphene-TMD heterostructures including many-body effects under a wide range of gate voltage at various temperatures without and with optical pumping.

**Acknowledgements**

This work was supported in part by the Ministry of Science and Technology (MOST), Taiwan under contract nos. 107-2112-M-001-032 and 108-2112-M-001-041.